\documentclass[article]{aa}
\usepackage{orcidlink}
\usepackage{graphicx}
\usepackage{txfonts}
\usepackage{amsmath,amssymb}
\usepackage{microtype}
\usepackage{hyperref}
\usepackage[nameinlink,capitalize]{cleveref}
\usepackage{csquotes}
\usepackage[switch]{lineno}

\newcommand{\kms}{$\mathrm{km\ s^{-1}}$}

\begin{document}

\title{X-rays from shock-heated gas in recurrent-nova remnants}
\subtitle{Nested nova shells in a structured circumstellar medium}

\author{
  Sergio Mart\'{\i}nez-Gonz\'alez
  \thanks{Corresponding author. E-mail: sergiomtz@inaoep.mx}%
  \inst{1}\orcidlink{0000-0002-4371-3823}
  \and
  Santiago Jim\'enez
  \inst{2}\orcidlink{0000-0003-2808-3146}
  \and
  Richard W\" unsch
  \inst{2}\orcidlink{0000-0003-1848-8967}
  \and
  Tiina Liimets
  \inst{3}\orcidlink{0000-0003-2196-9091}
}

\institute{
  Instituto Nacional de Astrof\'\i sica, \'Optica y Electr\'onica, AP 51, 72000 Puebla, M\'exico
\and
  Astronomical Institute of the Czech Academy of Sciences,  Bo\v{c}n\'\i\ II 1401/1, 141 00 Praha 4, Czech Republic
\and
  Tartu Observatory, University of Tartu, Observatooriumi 1, T\~oravere 61602, Estonia
}

\abstract
{Recurrent symbiotic novae such as RS~Oph show extended X-ray emission associated with the interaction of nova ejecta with the circumstellar environment.}
{We investigate how repeated eruptions over more than a century structure the circumstellar medium and govern the long-term X-ray evolution of a symbiotic recurrent nova.}
{We perform three-dimensional hydrodynamical simulations for a fiducial nine-eruption sequence spanning 130 years, modeling each nova as a supersonic bipolar shell expanding into a wind-shaped circumstellar medium, and post-process the resulting density and temperature distributions to compute X-ray emission in the 0.5--2.0 and 2.0--10.0~keV bands.}
{The eruption sequence excavates a bipolar cavity bounded by nested shells: soft X-rays trace dense compressed interfaces, whereas hard X-rays arise from the hottest shocked gas, including shell rims and the excavated nova remnant interior. As the remnant expands, the soft-band emission shows a gradual long-term decline associated with the decreasing emission measure, while the hard band evolves episodically, with individual eruptions imprinting distinct flares whose timing and relative strength change as the shell system grows. The extended X-ray morphology becomes progressively smoother and more volume-filling as density contrasts between successive shells are reduced.}
{Our simulations identify extended shell--shock emission as a natural consequence of the nested-shell evolution in recurrent novae: remnant expansion regulates the slow fading of the soft component, and renewed ejecta–shell encounters drive the hard-band variability on decade-long timescales. For a fiducial recurrent nova, the predicted diffuse soft X-ray luminosities are broadly consistent with the order of magnitude inferred for RS~Oph, indicating that shock-heated gas can contribute appreciably to the observed X-ray environment.}

\maketitle

\section{Introduction}
\label{intro}

Recurrent novae, exemplified by systems such as RS~Ophiuchi (RS~Oph) and T Coronae Borealis (T~CrB), serve as important probes of shock physics in extreme astrophysical environments \citep{Zhengetal2024}. These systems undergo multiple outbursts over time, during which a white dwarf in a close binary ejects material at velocities ranging from several hundred to thousands of \kms \citep[e.g.][]{Hernanz2008}. Among symbiotic recurrent novae, in which a white dwarf accretes from the wind of a red-giant companion, RS~Oph offers the best-observed eruption record, with documented outbursts in 1898, 1907, 1933, 1945, 1958, 1967, 1985, 2006 \citep[][and references therein]{Schaefer2010}, and more recently in 2021 \citep{Pageetal2022}. As a result, high-velocity ejecta interact with a dense circumstellar medium (CSM), often dominated by the red giant wind in symbiotic systems \citep{MooreBildsten2012} as well as with the remnants of previous outbursts \citep{Tappertetal2020}. These interactions span the immediate post-outburst shocks on timescales of days to weeks and the secular restructuring of the CSM over many recurrence cycles.

Multi-wavelength observations have revealed the complex structure of these interactions. For example, extended X-ray emission detected in RS~Oph highlights the presence of a bipolar outflow and shock-heated plasma expanding at velocities up to $\sim4600$ km~s$^{-1}$ \citep{MontezJretal2022,Zhengetal2024}. A similar smaller-scale bipolar structure has been observed in the optical range with the Hubble Space Telescope (HST) following the 2006 outburst (\citealt{Bodeetal2007,2009ApJ...703.1955R}). In addition, a super-remnant around RS~Oph has been observed to have a diameter of 70 pc, indicating that the system may have experienced many more nova events than the nine events recorded during recent history \citep{Sharaetal2025}. Comparable bipolar or strongly axisymmetric outflows have also been inferred in other symbiotic recurrent novae, including V745~Sco and V3890~Sgr, from X-ray and infrared emission-line diagnostics \citep{Orlandoetal2017,Evansetal2022}. The long-term structure produced by recurrent eruptions is also relevant for the eventual fate of these systems, since symbiotic recurrent novae have been discussed as potential Type Ia supernova progenitors, whose shocks would subsequently expand through the circumstellar environment assembled by previous eruptions \citep[e.g.][]{MooreBildsten2012,SerranoHernandezetal2025}.
 
The optical emission-line luminosities of nova shells commonly show an early shallow stage followed by fading as the shell expands and densities drop; very old shells may flatten at late times \citep{Tappertetal2020}. Recurrent novae deviate from this general behavior, plausibly because their ejecta interact with denser material, including shocks between ejecta from different outbursts or remnants from earlier eruptions \citep{Tappertetal2020,Downesetal2001}. This remnant memory makes the accumulated shell system central to the long-term shock evolution across many eruptions.

\citet{Bode1985} pioneered the modeling of RS~Ophiuchi's 1985 outburst by analyzing the interaction between nova ejecta and the red giant's wind. Utilizing X-ray data from EXOSAT at 55 days post-outburst, they demonstrated that RS~Oph evolves similarly to a supernova remnant on much shorter timescales. Early post-outburst studies probe the immediate shocked environment of the binary on day-to-month timescales, with X-ray luminosities reaching $\gtrsim 10^{35}\ {\rm erg\ s^{-1}}$, and, once the supersoft source emerges, residual hydrogen burning on the white dwarf surface \citep[e.g.][]{Bodeetal2006,Orioetal2023}. Detailed three-dimensional calculations of individual eruptions in RS~Oph, U~Sco, and T~CrB follow this compact post-outburst interaction on binary and circumbinary scales \citep{Orlandoetal2009,DrakeOrlando2010,Orlandoetal2025}.

Complementing these early post-outburst studies, we use three-dimensional hydrodynamic simulations to investigate the weaker, extended X-ray emission generated by repeated ejecta--wind and ejecta--shell interactions over many nova cycles. These interactions build a nested-shell remnant that emits diffuse X-rays between outbursts. Late-time \textit{Chandra} observations of RS~Oph infer unabsorbed luminosities of order a few $10^{31}\ {\rm erg\ s^{-1}}$ for this extended component \citep{MontezJretal2022}.

The manuscript is structured as follows. Section \ref{sec:model} describes the hydrodynamical scheme, the initial conditions for the stellar wind and the multiple eruption events (Section \ref{subsec:hydro}), and the post-processing of the hydrodynamical results to obtain the soft and hard X-ray fluxes (Section \ref{subsec:post-process}). In Section \ref{results}, we discuss the morphological and X-ray luminosity evolution, while the summary and our main conclusions are presented in Section \ref{conclusions}.

\section{Numerical Setup}
\label{sec:model}

Our calculations follow the evolution of the extended shock-heated ejecta and circumstellar remnant produced by repeated nova cycles. The central binary source, including X-rays from the accretion disc, boundary layer, wind mass transfer, and the short-lived supersoft emission powered by residual hydrogen burning on the white dwarf surface, is not included \citep{Boothetal2016}.\footnote{This supersoft component originates in the unresolved central-source and immediate post-outburst region, on AU or sub-AU scales and day-to-week timescales. In the present calculations the finest cell size is $\sim 80$ AU; treating this phase self-consistently would require either a much higher  resolution across the full remnant or a separate sub-grid prescription.} The resulting X-ray emission is therefore that of the resolved ejecta--CSM interaction.

\subsection{Hydrodynamical Scheme}
\label{subsec:hydro}

We perform three-dimensional hydrodynamical calculations with the adaptive-mesh-refinement (AMR) code \texttt{FLASH}\,4.6 \citep{Fryxelletal2000}. Our setup follows the nested‐shell prescription of \citet{SerranoHernandezetal2025}, originally devised for supernova remnants interacting with pre-existent eruption shells and here adapted to recurrent novae, with progenitor‐dependent winds and eruptions implemented via the \textsc{Wind} module \citep{Wunschetal2008,Wunschetal2017,SerranoHernandezetal2025}. The red-giant wind is used to establish the initial circumstellar density and velocity profiles before the first eruption: \begin{equation} \rho_{\rm w}(r)=\frac{\dot M}{4\pi\,v(r)\,r^{2}}, \qquad v(r)=v_\infty\,\frac{r}{R_{\rm w}}. \end{equation} This prescription ensures that the flow reaches its terminal speed $v_\infty$ at $r=R_{\rm w}$ and then streams freely with $v=v_\infty$ outside. In these equations, $\dot{M}$ is the stellar mass-loss rate, and $r$ is the distance to the source. We adopt an initial red-giant wind with $\dot M=10^{-7}\,M_\odot\,\mathrm{yr^{-1}}$ and $v_\infty=20$~\kms, consistent with previous symbiotic-nova calculations \citep[e.g.][]{Walderetal2008}. The wind has temperature $T_{\rm w}=10^{4}\,\mathrm{K}$, and the inner scale of the prescribed profile is $R_{\rm w}=0.008$ pc. The wind profile is imposed over the computational domain before the eruption sequence starts, with no subsequent wind mass or energy injection. No dust is included in the wind material.

\begin{figure}[ht]
\centering
\includegraphics[width=1.0\columnwidth]{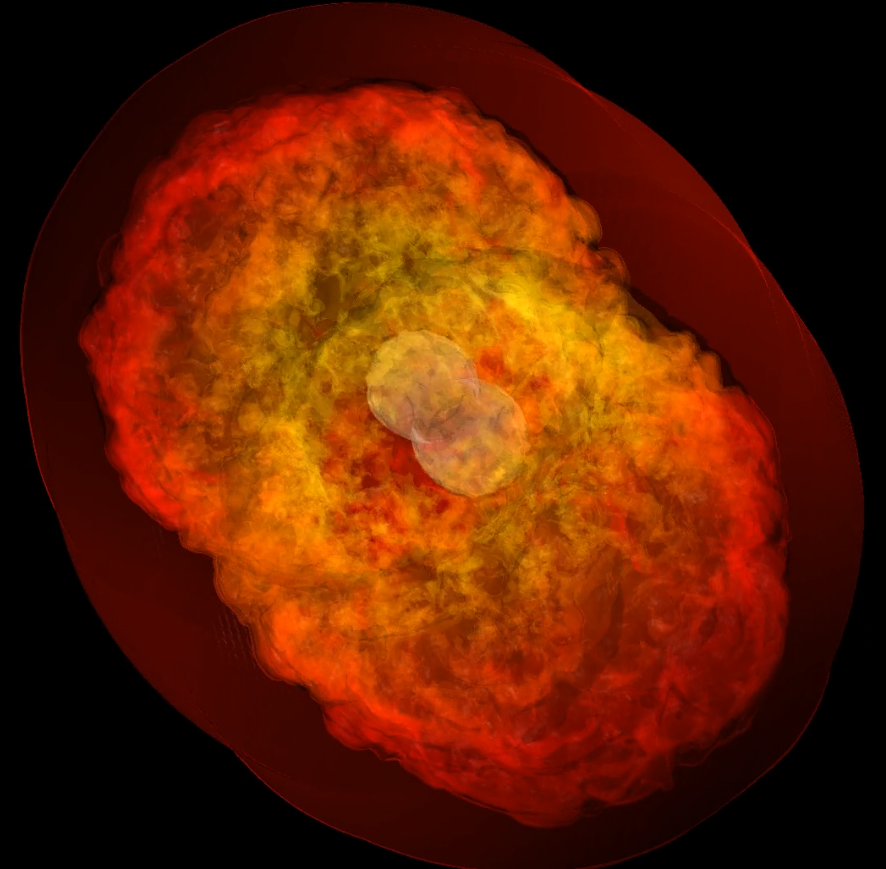}
\caption{Projection of nested shells in our hydrodynamical modeling of a fiducial RS~Oph-like recurrent nova. The image shows the density structure of the remnant one year after one of the simulated nova events over the full computational domain ($0.2\times0.2\times0.2$ pc$^{3}$); taking into account the distance of 2.4~kpc, this translates into $17''.5\times17''.5\times17''.5$ in the plane of the sky; the polygonal mesh representation is used for visualization purposes only. Note the inner bipolar structure; the color scheme is purely illustrative.}
\label{fig:RSOph_RNe}
\end{figure}

Each nova is introduced with a mass $M_{\rm ej}$ and kinetic energy $E_k$, deposited at an initial radius of $R_{\rm e} = 2.77 \times 10^{16}\,\mathrm{cm}$ ($0.009\,\mathrm{pc}$). The ejecta are inserted as a thin, hollow spherical shell with inner radius $0.95\,R_{\rm e}$ and outer radius $R_{\rm e}$ \citep[see][for details]{SerranoHernandezetal2025}, with clumpy inhomogeneities seeded through white-noise density perturbations \citep{MartinezGonzalezetal2018}. As the ejecta are left to evolve, the bipolar morphology is achieved via a latitude-dependent expansion velocity (see Figure \ref{fig:RSOph_RNe}):

\begin{equation}
v_{\textrm{e}} = \frac{r}{R_\textrm{e}}\,  
\left[ 1 - \alpha \left( \frac{1 - e^{-2\beta \sin^2 \varphi}}{1 - e^{-2\beta}} \right) \right]v_{\textrm{max,e}}
\end{equation}

giving the radial and angular dependence of the expanding ejecta \citep{Franketal1995,Blondin1995,Smith2006,SerranoHernandezetal2025}. The term $v_{\textrm{max,e}}$ in the equation is the maximum expansion velocity of the ejecta, and is calculated from energy conservation from the initial ejecta mass and energy. The parameters $\alpha$ and $\beta$ determine the pole-to-equator velocity contrast, and the overall morphology of the erupted ejecta, respectively, while $\varphi$ is the polar angle (see Table \ref{tab:sim_params} and \citet{SerranoHernandezetal2025} for further details).

We adopt as fiducial cadence template the recurrence intervals implied by the documented RS~Oph eruptions, $\Delta t = 9,\,26,\,12,\,13,\,9,\,18,\,21,$ and $15$ years \citep{Schaefer2010,Pageetal2022}. The first simulated eruption is placed at $t=0$ years, and subsequent eruptions are labeled by elapsed time and sequence number. Based on previous studies \citep[e.g.][and references therein]{Zhengetal2024,MooreBildsten2012}, we adopt $M_{\rm ej} = 2\times 10^{-6}$ M$_\odot$ and $E_k = 4.02\times10^{44}$~erg for each eruption and assume the formation of $2\times10^{-8}$ $M_\odot$ of dust per eruption (dust‐to‐gas ratio $10^{-2}$) following \citet{Banerjeeetal2023}. For homogeneous ejecta, this corresponds to an expansion velocity of approximately 4600 \kms, representative of fast RS~Oph-like symbiotic recurrent novae \citep{Bodeetal2007,Tomovetal2023,Zhengetal2024}.

Radiative cooling at solar metallicity is implemented with the tabulated optically thin collisional-ionization-equilibrium (CIE) curve of \citet{Schureetal2009}. We do not evolve ionization fractions in the hydrodynamics, so the cooling follows CIE even in regions where the plasma is out of ionization equilibrium. In addition, dust-induced cooling is computed on-the-fly with the \textsc{Cinder} module \citep{MartinezGonzalezetal2018,MartinezGonzalezetal2019,MartinezGonzalezetal2022}. The ejecta adopt a dust-to-gas mass ratio of $10^{-2}$, and grain sizes are drawn from a log-normal distribution, $n(a)\propto a^{-1}\exp\left[-\frac{1}{2}\left(\ln(a/a_{0})/\sigma\right)^{2}\right]$, characterized by $a_{0} = 0.1\ \mu\mathrm{m} \text{ and } \sigma = 0.7, \text{ truncated at } a_{\min} = 0.005\ \mu\mathrm{m} \text{ and } a_{\max} = 0.5\ \mu\mathrm{m}$.

The computational domain is a cubic box, centered at the origin of the Cartesian coordinate system, spanning ($0.2$ pc)$^3$. To accurately capture the evolving shock structures and their interactions, we employ AMR with a refinement threshold of $\rho = 5\times10^{-20}$ g cm$^{-3}$, a minimum cell size $\Delta x_{\min}=3.9\times10^{-4}$ pc, a maximum cell size $\Delta x_{\max}=7.8\times10^{-4}$ pc, and outflow boundary conditions.

\subsection{X-ray post-processing}
\label{subsec:post-process}

We post-process each hydrodynamical output with \textsc{XSPEC v.12.12} \citep{Arnaud1996} to obtain band-integrated fluxes in the \textit{soft} (0.5--2.0~keV) and \textit{hard} (2.0--10.0~keV) ranges. From the grid, the local mass density and temperature in each cell are read and converted to the electron number density, $n_{\mathrm e}$, and the plasma temperature $kT$ (in keV). A base abundance pattern from \citet{AndersandGrevesse1989}, normalized to hydrogen at $Z_\odot$, is modified using the elemental enhancement and depletion factors reported by \citet{Orioetal2023} for RS~Oph.

X-ray emissivities are evaluated throughout the full 3-D computational domain, cell by cell, with the non-equilibrium ionization (NEI) plane-parallel shock model \texttt{vvpshock} in \textsc{XSPEC} \citep{Borkowskietal2001}, using NEI atomic data from \textsc{AtomDB} \citep{Smithetal2001}. NEI is applied only in post-processing, whereas the gas-phase radiative cooling in the hydrodynamical evolution uses the CIE cooling function described in Section~\ref{subsec:hydro}. For the dust-to-gas mass ratio adopted here, $\mathcal{D}=10^{-2}$, gas--grain collisions dominate the cooling at $T\gtrsim2\times10^{6}$~K and can reduce the cooling time by up to about two orders of magnitude relative to gas-phase cooling alone (see Figure~1 of \citealt{MartinezGonzalez2025b}). The large-scale secular evolution is therefore governed mainly by a cooling channel that is weakly sensitive to the ionic charge-state distribution, while the CIE approximation primarily affects the detailed thermal and spectral structure. For the X-ray synthesis we adopt $kT_{\rm e}=\beta_{\rm e}kT$ with $\beta_{\rm e}=1$, so that the electron temperature is identified with the hydrodynamic temperature, and take the electron density to be $n_{\rm e}=\eta n_{\rm H}$ with $\eta=1.2$ at the relevant temperature range \citep{Schureetal2009}, where $n_{\rm H}$ is the local hydrogen number density. These quantities are provided to \texttt{vvpshock} as input plasma parameters; the model then computes the ionic charge-state distribution internally by solving the ionization and recombination equations for the specified $kT_{\rm e}$ and ionization-timescale range \citep{Borkowskietal2001,Smithetal2001}. The plane-parallel geometry of \texttt{vvpshock} assumes an idealized shock history, so mixed or repeatedly shocked cells may have more complex ionization histories. The ionization-equilibration timescales of the diffuse remnant exceed the duration of the simulated evolution by several orders of magnitude, supporting the adopted NEI treatment \citep{SmithandHughes2010}.

To assign the NEI timescales in a recurrent-eruption flow, we separate newly shocked ejecta from older mixed ejecta with a pressure mask applied over the full cube. We use an empirically calibrated pressure threshold to identify cells occupied by the most recently ejected material, which lies immediately behind the latest shock front, and classify cells with thermal pressures above $3.16\times10^{-7}$ dyn cm$^{-2}$ as the freshest ejecta (see Appendix \ref{app:pp}). For these cells we set the characteristic time to the time elapsed since the last eruption. All other cells are treated as material from earlier eruptions that has already interacted with the composite shell and mixed efficiently, and we assign them a single representative characteristic time equal to the mean time elapsed since all prior eruptions at that time. The \texttt{vvpshock} ionization-timescale range is then set locally to $\tau_{\max}=n_{\rm e}t_{\rm char}$ and $\tau_{\min}=0.1\tau_{\max}$, in units of s cm$^{-3}$ \citep{Borkowskietal2001}, with both values restricted to the range adopted in our post-processing and to the bounds allowed by \texttt{vvpshock}. Accordingly, the pressure-selected cells use a short NEI clock, while all other cells use the longer mixed characteristic timescale.

We use the standard \textsc{XSPEC} normalization for each cell, with $z=0$ and the hydrogen number density equal to the ion number density. The soft- and hard-band cell fluxes are summed over the computational domain at each epoch. The reported luminosities are intrinsic (unabsorbed), with no radiative-transfer calculation. Line-of-sight attenuation is evaluated separately to quantify its magnitude and viewing-angle dependence.

\begin{figure*}[ht]
\centering
\includegraphics[width=2.2\columnwidth]{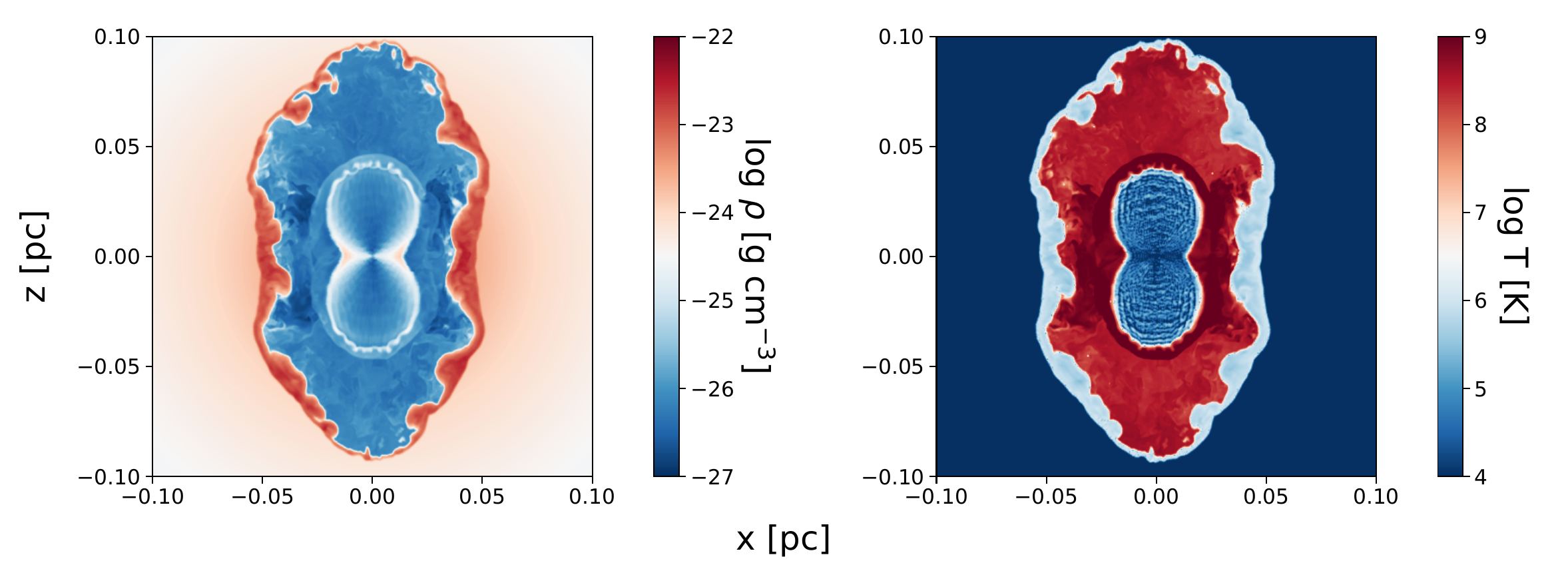}
\caption{Two-dimensional cuts of the mass density (left) and temperature (right) distributions in our fiducial symbiotic recurrent nova model after 110 years. Note that the lower lobule is slightly larger than the upper one. This asymmetry arises from interactions with the inhomogeneous ejecta during the early stages, as random clumps (introduced as white noise) affect each lobule differently. These slight initial differences become more apparent over time.}
\label{fig:RSOph_map}
\end{figure*}

\begin{table*}[ht]
\centering
\begin{tabular}{l l l l}
\hline\hline
Parameter                        & Value / Range                                           & Units           & Description \\
\hline
Initial wind mass‐loss rate              & $10^{-7}$                                        & $M_\odot\,\mathrm{yr^{-1}}$ & Initial red‐giant wind ($\dot{M}$) \\
Initial wind terminal velocity           & 20                                                     & $\mathrm{km\,s^{-1}}$  & Initial red‐giant wind ($v_w$) \\
Eruption mass                    & $2\times10^{-6}$  & $M_\odot$  & Shell mass per nova ejecta ($M_{\rm ej}$) \\
Eruption kinetic energy          & $4.02\times10^{44}$  & erg           & Kinetic energy per nova event ($E_k$) \\
Ejecta injection radius          & 0.009                                                   & pc              & Initial ejecta radius ($R_{\rm e}$) \\
Ejecta dust mass                 & $2\times10^{-8}$                                        & $M_\odot$       & Dust mass per nova ejecta \\
Shell shape parameters           & $\alpha = 0.9$, $\beta = 0.1$                           & —               & Shape parameters in density profile \\
Elemental abundances             & Solar $\times$ RS~Oph factors & — & \citep{AndersandGrevesse1989,Orioetal2023} \\
Domain size                      & $0.2\times0.2\times0.2$                                  & pc$^3$          & Cubic computational domain \\
AMR resolution                   & $\Delta x_{\min} = 3.9\times10^{-4}$; $\Delta x_{\max} = 7.8\times10^{-4}$ & pc & Minimum/maximum cell size \\
Simulation period                & 130                                               & years              & Range of simulated recurrent nova events \\
Simulated outburst epochs
& $0,\,9,\,35,\,47,\,60,\,69,\,87,\,108,$ and $123$
& years
& Epochs relative to the start of the simulation \\
Emission model                   & \texttt{vvpshock}                                         & —               & NEI model \citep{Borkowskietal2001}\\
Energy bands                     & 0.5--2.0 (soft); 2.0--10.0 (hard)                           & keV             & Bands for flux integration \\    
Adopted distance                 & 2.4
                                 & kpc             & RS~Oph-based
\citep{Bailer-Jonesetal2021}\\
\hline
\end{tabular}
\caption{Parameters used in the hydrodynamical simulations and post-processing of the fiducial symbiotic recurrent nova model. The table summarizes the physical assumptions for the wind, nova ejecta, dust content, and domain setup, along with the spectral modeling configuration used to compute the X-ray emission.}
\label{tab:sim_params}
\end{table*}

\begin{figure*}[ht]
\centering
\includegraphics[width=2.2\columnwidth]{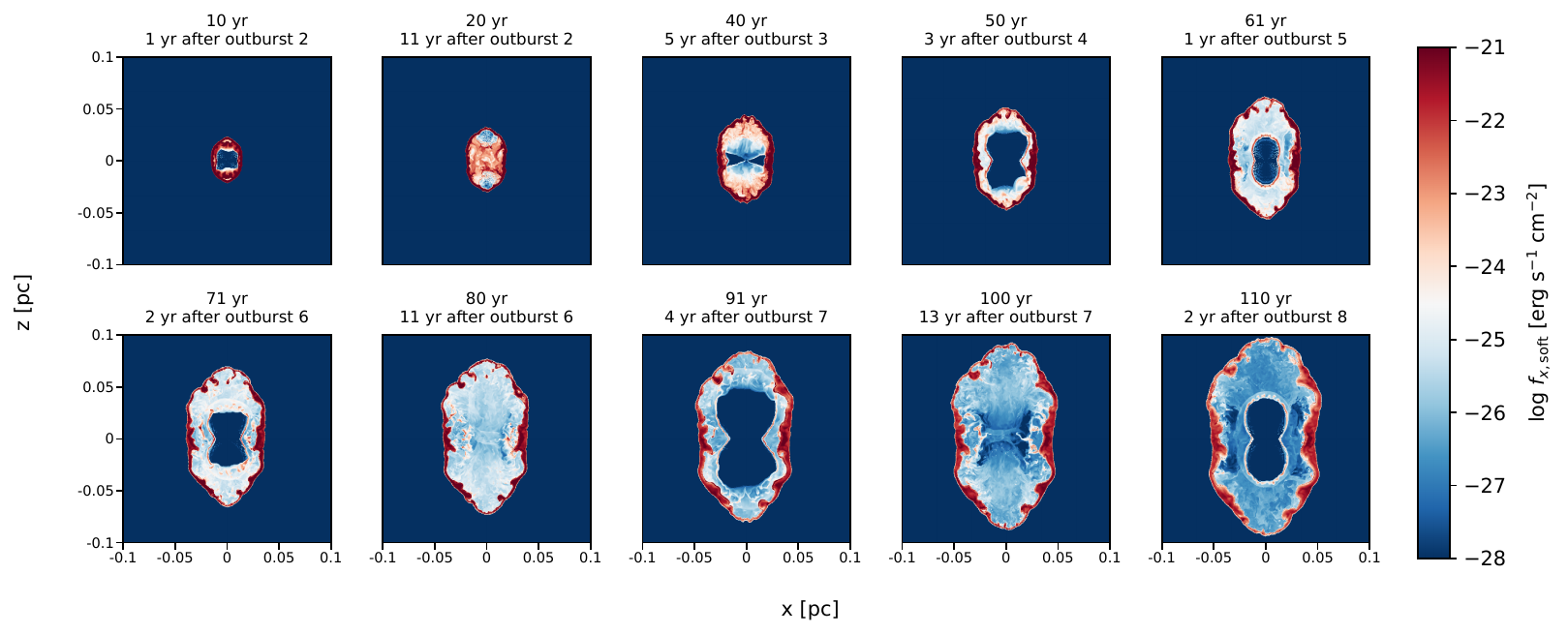}
\caption{Two-dimensional cuts showing the evolutionary sequence of the soft X-ray flux distribution in our fiducial symbiotic recurrent nova model. Each cut has a thickness of one cell, after remapping the AMR data onto a uniform mesh. The assumed distance is $2.4\ \mathrm{kpc}$. The panels show epochs at $t=10,\,20,\,40,\,50,$ and $61$ years (top row) and $t=71,\,80,\,91,\,100,$ and $110$ years (bottom row). The annotation above each panel also gives the elapsed time since the most recent outburst and the cumulative number of outbursts at that epoch. Bright soft X-ray rims delineate dense shells formed by accumulation of nova ejecta. Internal low-emission cavities and shell--shell interfaces develop as successive eruptions reshape the remnant.}
\label{fig:RSOph_soft}
\end{figure*}

\begin{figure*}[ht]
\centering
\includegraphics[width=2.2\columnwidth]{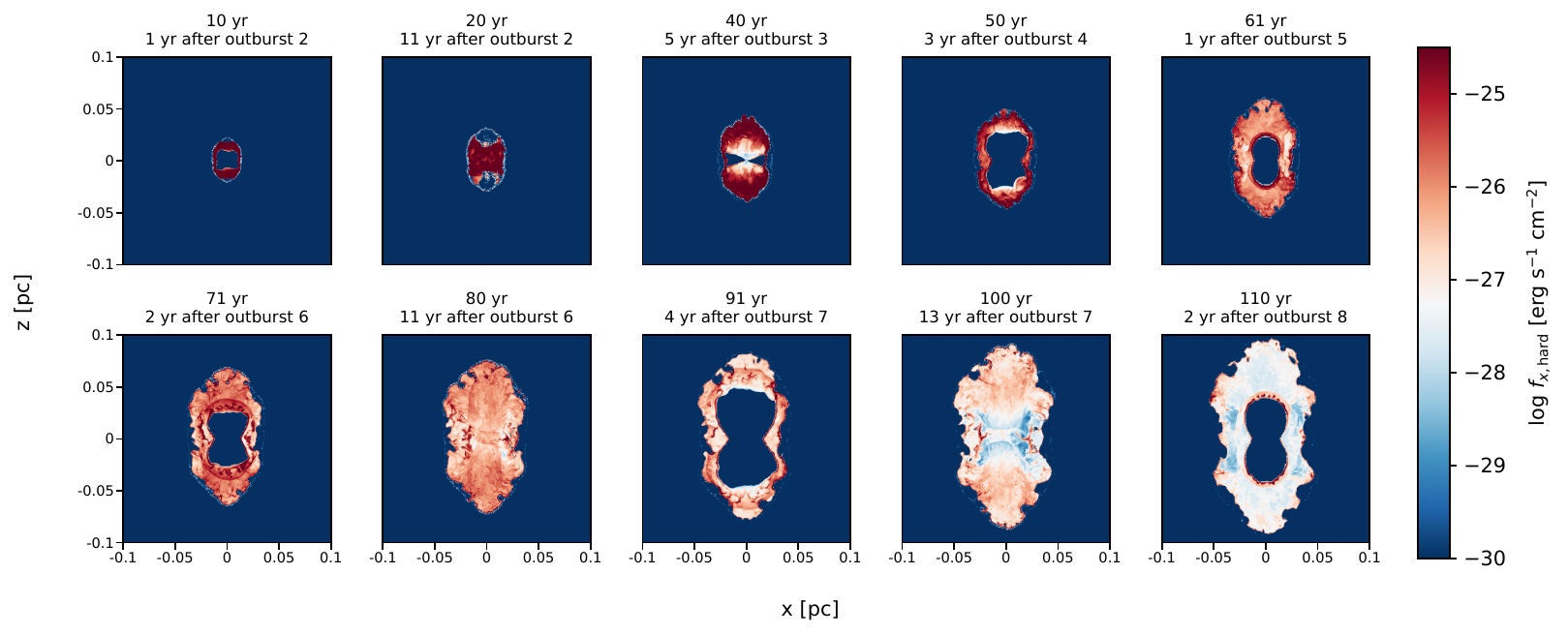}
\caption{Same as in Figure~\ref{fig:RSOph_soft}, but for the evolution of the hard X-ray (2.0--10.0~keV) flux distribution. Hard X-ray hotspots trace the hottest shocked gas along the composite shell rims and interior interaction fronts. At later times, the emission extends through the remnant interior as temperatures become more uniform and density contrasts weaken.}
\label{fig:RSOph_hard}
\end{figure*}

\section{Results} 
\label{results}

\subsection{Morphological Evolution and X-ray Maps}

Figure~\ref{fig:RSOph_map} displays 2-D cuts along the $x-z$ plane (remapped from the full-resolution AMR output onto a uniform grid) for the fiducial model at $t=110$ years after the first simulated eruption. The gas mass density (left panel) shows the excavated bipolar cavity created by previously ejected shells. The outburst at $t=108$ years produces an inner bipolar structure with a physical extent of 0.058 pc, corresponding to an angular size of $\sim5''$ in the plane of the sky at a distance of 2.4 kpc. This scale is broadly compatible with the resolved X-ray structure seen in RS~Oph (see Figure 1 of \citealt{MontezJretal2022}), which provides a well-observed reference point for this class, although our model is not tailored to reproduce RS~Oph epoch by epoch. The right panel illustrates shock-heated gas that attains temperatures of up to $\sim 10^8\ \mathrm{K}$ within the cavity. 

\begin{figure*}[ht]
\centering
\includegraphics[width=2.0\columnwidth]{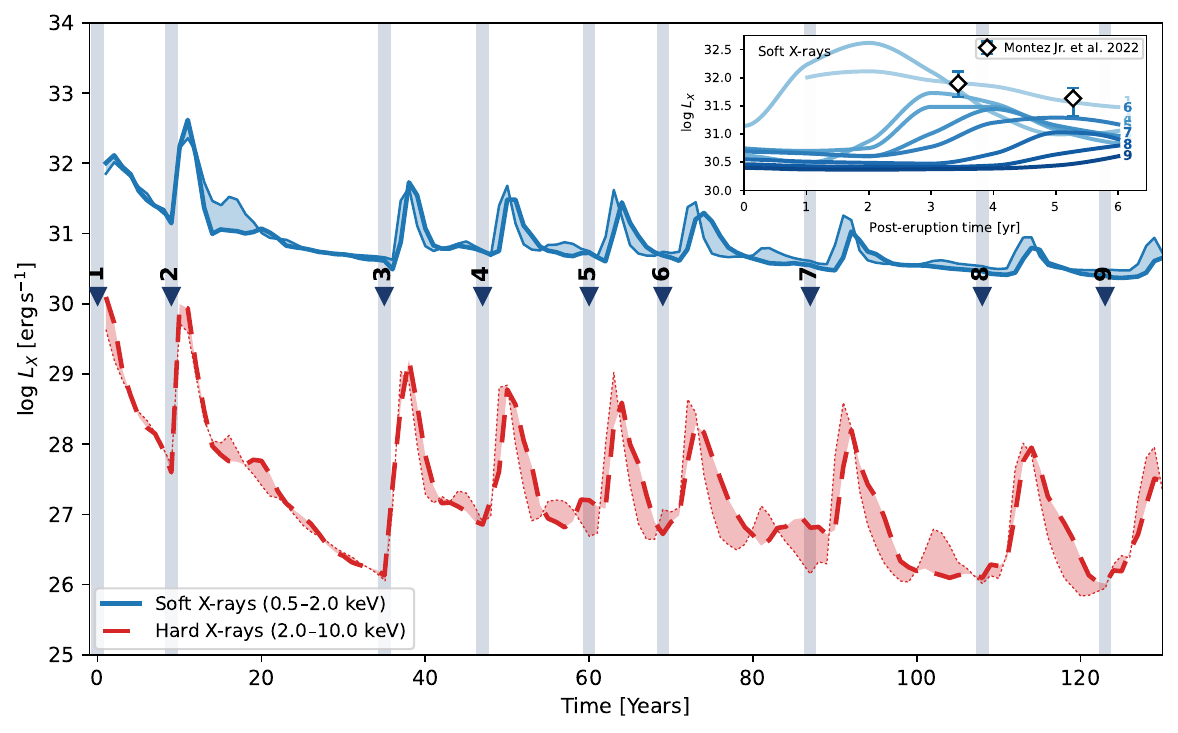}
\caption{Post-processed X-ray luminosities for the fiducial recurrent nova eruption sequence over 130 years, measured from the first simulated eruption at $t=0$, in the soft (0.5--2.0~keV; solid blue) and hard (2.0--10.0~keV; dashed red) bands. The shaded bands show the difference between the fiducial run and the lower-refinement run. Vertical gray shaded stripes denote the eruption times, and the numbers identify the simulated eruption sequence. The inset shows the simulated 0.5--2.0~keV luminosity during the first six years after each eruption, with the blue curves labeled by eruption number. Diamond markers show the extended 0.5--1.8~keV luminosities inferred for RS~Oph by \citet{MontezJretal2022} at 1254 and 1927 days after the 2006 eruption.}
\label{fig:Lx}
\end{figure*}

\subsection{Evolution in Soft and Hard X-rays}
\label{subsec:evol}

Figure~\ref{fig:RSOph_soft} presents the 2-D $x-z$ cuts of the soft X-ray evolutionary sequence (0.5--2.0~keV), while Figure~\ref{fig:RSOph_hard} shows the 2-D $x-z$ cuts of the hard X-ray evolutionary sequence (2.0--10.0~keV), both spanning $t=10$--110 years after the first simulated eruption. The panels correspond to $t=10,\,20,\,40,\,50,\,61,\,71,\,80,\,91,\,100,$ and $110$ years. In the soft band, the emission is dominated by dense shell rims, with the brightest regions tracing the composite shell interfaces produced by shocks interacting with earlier ejecta, while low-emission cavities persist within the nested-shell structure.

In the hard band, the emission traces the hottest shocked gas along the composite shell rims and interior interaction fronts. At later times, it extends through the remnant interior as temperatures become more uniform and density contrasts weaken.

This evolution is qualitatively similar to the post-outburst fading of an initially sharp bipolar structure into a more diffuse morphology in RS~Oph (see Figure 1 in \citealt{MontezJretal2022}). In these data, the initially well-defined bipolar morphology fades and becomes more diffuse over the next few years.
The same evolutionary pattern, although on a smaller physical scale, is also evident in the optical HST data from the 2006 outburst. Indeed, the initially clear bipolar morphology detected shortly after the outburst in 2006 \citep{Bodeetal2007} is considerably less pronounced after one year \citep{2009ApJ...703.1955R}.

Figure~\ref{fig:Lx} shows our simulated X-ray luminosities (integrated over the full computational volume) in the soft (0.5--2.0~keV) and hard (2.0--10.0~keV) bands over a 130-year span following the first eruption at $t=0$. The soft X-ray luminosity (blue curve) rises rapidly to $4.16\times10^{32}\ \mathrm{erg\ s^{-1}}$ at $t\simeq11$ years and then declines secularly; after $t\simeq22$ years it varies between $2.33\times10^{30}$ and $5.36\times10^{31}\ \mathrm{erg\ s^{-1}}$. From the early maximum to the final sampled epoch at $t\simeq128$ years the soft-band luminosity declines by $\simeq2.18$~dex, with subsequent eruptions producing short-lived enhancements.

In contrast, the hard X-ray luminosity (red curve) evolves episodically, with peaks occurring within a few years of eruptions. The peak values decline from $1.25\times10^{30}\ \mathrm{erg\ s^{-1}}$ after the first eruption to $8.93\times10^{27}\ \mathrm{erg\ s^{-1}}$ after the $t=108$-year eruption. Following the $t=123$-year eruption, the hard-band luminosity reaches $1.36\times10^{27}\ \mathrm{erg\ s^{-1}}$ by $t\simeq128$ years. Values as low as $\sim10^{26}\ \mathrm{erg\ s^{-1}}$ occur between the major enhancements.

The inset shows the soft-band evolution of the individual cycles as a function of post-eruption time. The spread among the numbered curves arises from the different circumstellar structures established by the preceding eruptions. The extended 0.5--1.8~keV luminosities inferred by \citet{MontezJretal2022}, $7.9\times10^{31}$ and $4.3\times10^{31}\ \mathrm{erg\ s^{-1}}$, correspond to 3.43 and 5.28 years after the 2006 eruption. The decrease between these epochs is closely followed by the brightest simulated cycles, which reach a comparable luminosity scale and show a similar decline rate. The remaining differences likely reflect uncertainties in the ejecta properties and the density structure of the circumstellar medium, as well as our neglect of magnetic fields and anisotropic thermal conduction \citep[e.g.][]{Orlandoetal2008}. Moreover, we follow only nine cycles, whereas many symbiotic recurrent novae likely experience many more cycles, as suggested by the reported $\sim70$ pc super-remnant around RS~Oph \citep{Sharaetal2025}.

The sequence of hard X-ray flares shows that localized high-density shells continue to sustain strong shocks throughout the secular evolution. The hard band traces the hottest plasma produced during these transient interactions. In the simulations, the 2.0--10.0~keV luminosity remains elevated for several years after an eruption before declining toward its inter-eruption level.

All maps shown here adopt a single viewing geometry (line of sight along the $y$-axis). We also computed band-integrated luminosities for viewing angles of $0^\circ$, $30^\circ$, $60^\circ$, and $90^\circ$, including photoelectric absorption with the \textsc{XSPEC} \texttt{tbabs} model. In the extended remnant modeled here, $N_{\rm H}\lesssim10^{19}\ {\rm cm^{-2}}$, giving only $\sim1\%$ attenuation at 0.5~keV. For context, early post-outburst X-ray spectral fits of the 2021 eruption of RS~Oph by \citet{Orioetal2023}, which probe compact AU-scale gas not included in our calculations, imply $N_{\rm H}\sim10^{22}\ {\rm cm^{-2}}$. Viewing-angle-dependent absorption may therefore be important in that compact region, but is negligible for the diffuse remnant considered here.

To assess numerical convergence, we repeated the calculation with a lower maximum refinement level while keeping the domain and base grid unchanged; the shaded regions in Figure~\ref{fig:Lx} show the differences between the two runs. The time-weighted mean $|\Delta{\rm dex}|$ is $\simeq0.1396$ in the soft band and $\simeq0.3338$ in the hard band. 

\section{Conclusions}
\label{conclusions}

We have presented three-dimensional hydrodynamical simulations of a fiducial recurrent nova eruption sequence, covering 130 years, and we have examined the resulting soft and hard X-ray signatures. The adopted cadence is guided by RS~Oph, which also provides a well-observed reference for comparison. Our main findings are as follows:

\begin{enumerate}
\item The repeated eruptions build a nested bipolar shell system in which soft X-rays trace dense shell interfaces and low-emission cavities, while hard X-rays arise from the hottest gas along shell rims and interior interaction fronts. As the remnant evolves, the hard-band emission extends through a progressively more uniform interior. The resulting density and temperature structure provides a physically evolved circumstellar environment for subsequent explosive events, including a possible Type Ia supernova whose shock propagation would be shaped by the accumulated shell system.
\item Episodic boosts in soft X-ray luminosity follow each eruption, but they are modest and do not erase the long-term behaviour. The soft-band luminosity reaches an early maximum of $\sim4\times10^{32}\ \mathrm{erg\ s^{-1}}$ and subsequently remains mostly between $\sim10^{30}$ and a few $10^{31}\ \mathrm{erg\ s^{-1}}$. By the end of the calculation it has declined by about two orders of magnitude, reflecting the decreasing emission measure as the accumulated remnant expands. 
\item Hard X-ray emission is strongly episodic. Its peak luminosity decreases from $\sim10^{30}\ \mathrm{erg\ s^{-1}}$ after the first eruption to $\sim10^{28}\ \mathrm{erg\ s^{-1}}$ during the later cycles, with inter-eruption values dropping to $\sim10^{26}\ \mathrm{erg\ s^{-1}}$. These enhancements trace transient interactions between fast ejecta and the accumulated shell system. 
\item For comparison, the extended soft X-ray luminosity of RS~Oph inferred from the 2009 and 2011 \textit{Chandra} observations analyzed by \citet{MontezJretal2022} decreases from approximately $7.9\times10^{31}$ to $4.3\times10^{31}\ \mathrm{erg\ s^{-1}}$. The brightest cycles in our fiducial sequence reach a comparable luminosity scale and show a similar decline rate.
\end{enumerate}
Within the limitations of the model, these results show that nested-shell evolution can sustain long-lived diffuse soft X-ray emission between recurrent-nova eruptions.

\begin{acknowledgements}
This work is dedicated to the memory of Sergiy Silich, whose scientific guidance, generosity, and friendship meant so much to many of us. The authors thank the anonymous referee for the valuable suggestions. The authors thankfully acknowledge the computer resources, technical expertise and support provided by the Laboratorio Nacional de Supercómputo del Sureste de México, SECIHTI member of the network of national laboratories. SJ acknowledges support by the Czech Ministry of Education, Youth and Sports, through the INTER-EXCELLENCE II program, project LUC24023, and by the institutional project RVO:67985815. The software used in this work was developed in part by the DOE NNSA- and DOE Office of Science-
supported Flash Center for Computational Science at the University of Chicago and the University of
Rochester. {\it Software:} FLASH v4.6.2 \citep{Fryxelletal2000}, NumPy \citet{Numpy}, \textsc{Wind} \citep{Wunschetal2017}, \textsc{Cinder} \citep{MartinezGonzalezetal2018}, Matplotlib \citep{Matplotlib}, SciPy \citep{SciPy}, h5py \citep{collette_python_hdf5_2014}, \textsc{XSPEC}  \citep{Arnaud1996}.
\end{acknowledgements}
\bibliographystyle{aa}
\bibliography{novae.bib}

\begingroup
\let\AAclearpage\clearpage
\let\clearpage\relax
\begin{appendix}
\let\clearpage\AAclearpage

\section{Pressure-based identification of freshly shocked ejecta}
\label{app:pp}
We identify the freshest ejecta using a pressure-based tracer intended to locate the immediate post-shock layer of the most recent eruption. During the propagation of the latest shock, the cells directly behind the shock front form a contiguous region of elevated thermal pressure. After the shock interacts with the composite shell, the characteristic post-shock pressures in that region drop rapidly as the flow decelerates and the shocked material is redistributed, and mixing then proceeds efficiently. This motivates using a pressure threshold as an empirical proxy for detecting the freshest post-shock ejecta.
We calibrate the threshold by tracking the pressure field in representative snapshots spanning multiple eruptions and by comparing it with the morphology of the newest shock. The pressure distribution typically exhibits a distinct high-pressure component associated with the layer immediately behind the latest shock. A threshold of $P_{\rm th}=3.16\times10^{-7}$ dyn cm$^{-2}$ isolates this component across the epochs used for X-ray synthesis while excluding the lower-pressure volume containing material processed by earlier eruptions.
We adopt this criterion solely to assign NEI clocks in post-processing. Cells with $P>P_{\rm th}$ are tagged as freshly shocked ejecta and assigned $t_{\rm char}=t-t_{\rm last}$, the time elapsed since the most recent eruption. All remaining cells are treated as older, previously processed material and assigned a single representative $t_{\rm char}$ given by the average elapsed time since all eruptions that had occurred by the snapshot. We verified that modest variations around $P_{\rm th}$ do not qualitatively change the identified morphology of the freshest post-shock layer.
\end{appendix}
\endgroup

\end{document}